\documentclass[12pt]{article}

\usepackage{epsfig}
\usepackage{amsmath}
\usepackage{amssymb}

\begin{document}

\marginparwidth 3cm
\setlength{\hoffset}{-1cm}
\newcommand{\mpar}[1]{{\marginpar{\hbadness10000%
                      \sloppy\hfuzz10pt\boldmath\bf\footnotesize#1}}%
                      \typeout{marginpar: #1}\ignorespaces}
\def\mda{\mpar{\hfil$\downarrow$\hfil}\ignorespaces}
\def\mua{\mpar{\hfil$\uparrow$\hfil}\ignorespaces}
\def\mla{\marginpar[\boldmath\hfil$\rightarrow$\hfil]%
                   {\boldmath\hfil$\leftarrow $\hfil}%
                   \typeout{marginpar:
                     $\leftrightarrow$}\ignorespaces}

\providecommand{\ads}{\text{AdS}}

\begin{titlepage}

\begin{flushright}
\today
\end{flushright}

\vspace{1cm}

\begin{center}
\baselineskip25pt
{\large\bf On the Flattening of Negative Curvature via T-Duality
           with a Non-Constant $B$-field}
\end{center}

\vspace{1cm}

\begin{center}
\baselineskip12pt
{Axel Krause\footnote{E-mail: {\tt krause@physics.umd.edu};\;
now at Department of Physics, University of Maryland, College
Park, MD 20742, USA}}
\vspace{1cm}

{\it Humboldt-Universit\"{a}t, Institut f\"{u}r Physik,
     D-10115 Berlin, Germany}

\vspace{0.3cm}
\end{center}

\vspace*{\fill}

\begin{abstract}
In an earlier paper, Alvarez, Alvarez-Gaum\'{e}, Barb\'{o}n and
Lozano pointed out, that the only way to ``flatten" negative
curvature by means of a T-duality is by introducing an
appropriate, non-constant NS-NS 2-form $B$. In this paper, we
are investigating this further and ask, whether it is possible
to T-dualize $\ads_d$ space to flat space with some suitably
chosen $B$. To answer this question, we derive the relation
between the original curvature tensor and the one of the
T-dualized metric involving the $B$-field. It turns out that
there is one component which is independent of $B$. By
inspection of this component we show that it is not possible to
dualize $\ads_d$ to flat space irrespective of the choice of
$B$. Finally, we examine the extension of $\ads$ to an $\ads_5
\times \text{S}^5$ geometry and propose a chain of S- and
T-dualities together with an $SL(2,\mathbb{Z})$ coordinate
transformation, leading to a dual D9-brane geometry.
\end{abstract}

\noindent
PACS: 11.25.-w \\
Keywords: T-Duality, NS-NS B-Field

\vspace*{\fill}

\end{titlepage}

By now the AdS-CFT conjecture \cite{Malda,Gubser,Witten} has
passed an enormous amount of tests (see \cite{Malda2} for a
review). Most of them explored the large $N$ duality between
the boundary-CFT and the supergravity on $\ads_5$. The common
radius of the $\text{S}^5$ and the length-scale of the
$\ads_5$, $R$, is given by $R^4 = (\alpha')^2 \lambda, $ with
$\lambda = g_{YM}^2 N$ representing the 't Hooft coupling. In
general supergravity as the low-energy limit of string-theory
is trustworthy only at large scales, i.e.~small curvatures.
Thus tests of the duality probing the supergravity regime
explore the $\lambda \rightarrow
\infty$ parameter region. This predicts how the CFT at large $N$
behaves in the extreme non-perturbative regime. \\
\phantom{a}\hspace{1em}
The interesting parameter regime, interpolating between the
perturbative $\lambda\rightarrow 0$ and the extreme
non-perturbative $\lambda \rightarrow \infty$ regime, demands
that we keep $\lambda$ finite. Since the closed string coupling
constant $g_s$ and the Yang-Mills coupling constant are related
via $g_{YM}^2 = 4\pi g_s$, the large $N$ limit with $\lambda$
finite, requires $g_s\rightarrow 0$. Thus via the AdS-CFT
conjecture, we are able to extract the full quantum information
about the CFT in the large $N$ limit by calculating simply IIB
string tree-diagrams. Unfortunately, this wonderful perspective
is obstructed by the fact, that IIB string-theory in the
RNS-formulation on an $\ads_5\times \text{S}^5$ background with
$N$
 units
of RR 5-form flux through the $\text{S}^5$ is still obscure
(see \cite{Polyakov} for an approach). On the other hand there
are proposals for a GS-formulation \cite{Tseytlin} which is
non-linear and therefore its quantization and computation of
string scattering amplitudes seems to be
difficult. \\
\phantom{a}\hspace{1em} Since string-theory on an $\ads_5\times \text{S}^5$
background is not readily available, one may seek resort to a
dual description of IIB string-theory on a, hopefully, easier
background. As a first step into this direction, we want to
analyze in this paper, whether pure $\ads_d$ space can be
dualized to flat Minkowski space. This would have important
consequences for the cosmological constant in string-theory
brane-worlds based on a bulk $\ads_d$ space \cite{K3}. The
concrete motivation for this problem stems from the
observation, made in \cite{Alvarez}, that the only way to
``flatten" negative curvature under T-duality is by introducing
an appropriate torsion, generated by $B_{\mu\nu}$, in the
initial space-time. This can be seen from the following
formula\footnote{The index $\iota$ represents the isometry
direction, whereas $\alpha,\beta,\hdots$ label the remaining
directions.}, relating the dual curvature scalar ${\tilde R}$
to the initial curvature scalar
\cite{Alvarez}
\begin{equation}
   {\tilde R} = R + 4\,\triangle \ln k +\frac{1}{k^2}H_{\iota\alpha\beta}
                H^{\iota\alpha\beta} -\frac{k^2}{4}F_{\alpha\beta}
                F^{\alpha\beta} \; .
      \label{CurvatureScalar}
\end{equation}
Here $A_\alpha = k_\alpha/k^2$, $k_\alpha = g_{\iota\alpha}$
with the associated field-strength
$F_{\alpha\beta}=\partial_\alpha A_\beta -\partial_\beta
A_\alpha$. Furthermore, $\triangle \ln k=\frac{1}{\sqrt{-g}}
\partial_\alpha \big( \sqrt{-g} g^{\alpha\beta} \partial_\beta \ln k \big)$
denotes the d'Alembertian with respect to the initial metric.
As usual, the torsion 3-form is the field strength\footnote{Our
conventions are:
                          \begin{equation*}
                          B = \frac{1}{2!} B_{\mu\nu}
                              dx^\mu \wedge dx^\nu \; ,
                             \qquad
                          H = \frac{1}{3} H_{\mu\nu\rho}
                              dx^\mu \wedge dx^\nu \wedge dx^\rho \; .
                          \end{equation*}}
$H=dB$ of the NS-NS 2-form $B_{\mu\nu}$. The Killing vector,
corresponding to the assumed translational isometry exhibited
by the initial space-time, is given by $k_\mu$ and its norm
defined by $k=\sqrt{g_{\mu\nu}k^\mu k^\nu}$. One could now try
to solve (\ref{CurvatureScalar}) for $R=-\Lambda$ and ${\tilde
R}=0$, with $\Lambda$ a positive constant. But this would
include any solution of the vacuum Einstein equations. To
decide whether we arrive at flat space after
T-dualization, we have to regard the Riemann curvature tensor. \\
\phantom{a}\hspace{1em} Let us start quite generally by assuming
some coordinate representation of a $d$-dimensional manifold,
given by\footnote{Indices $\lambda,\mu,\nu,\pi$ run from
$0,\hdots,d-1$,
            whereas indices $\alpha,\beta,\gamma,\delta,\epsilon$ run over
            $0,\hdots,d-2$.}
$x^\mu=(x^0,x^1,\hdots,x^{d-2},x^{d-1}) =
 (x^\alpha,x^\iota) \; ; \; \alpha=0,\hdots,d-2$.
Furthermore, the initial metric is supposed to be of the form
\begin{equation}
   ds^2 =   g_{\alpha\beta} dx^{\alpha} dx^{\beta}
          + g_{\iota\iota} dx^\iota dx^\iota  \; .
           \label{InitialMetric}
\end{equation}
From the Buscher rules for T-duality
\begin{equation*}
   {\tilde g}_{\iota\iota} = \frac{1}{g_{\iota\iota}}
                                                \; , \qquad
   {\tilde g}_{\iota\alpha} = \frac{B_{\iota\alpha}}{g_{\iota\iota}}
                                                \; , \qquad
   {\tilde g}_{\alpha\beta} =
   g_{\alpha\beta} - \frac{\left( g_{\iota\alpha}g_{\iota\beta}
   -B_{\iota\alpha}B_{\iota\beta} \right)}{g_{\iota\iota}} \; ,
\end{equation*}
we observe, that our choice of $g_{\iota\alpha}=0$ cannot be
compensated by an appropriate choice of some real-valued
$B_{\iota\alpha}$ (only an imaginary would suffice). The reason
why we set nevertheless $g_{\iota\alpha}=0$, from the outset,
is the following. According to (\ref{CurvatureScalar}), a
non-vanishing $g_{\iota\alpha}=k_\alpha$ tends to make the
curvature scalar of the dual metric more negative. Since we
start with a negatively curved $\ads_d$ and aim to bring it to
flat space, a
$k_\alpha \not= 0$ would obstruct our intention. \\
\phantom{a}\hspace{1em} We assume that the metric $g_{\mu\nu}(x^\alpha)$
and the NS-NS 2-form $B_{\mu\nu}(x^\alpha)$ do not depend on
$x^\iota$. These conditions are exactly those, which are
required in order to leave the $\sigma$-model action
\begin{equation*}
   S=\frac{1}{2\pi}\int d^2 z \left( g_{\mu\nu} + B_{\mu\nu}
                              \right)
     \partial X^\mu {\bar \partial} X^\nu
\end{equation*}
invariant under infinitesimal shifts $x^\mu \rightarrow x^\mu +
\epsilon k^\mu$. The Killing vector associated with the
resulting abelian translational isometry, is $k^\mu
\partial_\mu = \partial/ \partial x^\iota$, with $k^\mu =
(k^\alpha, k^\iota) = (0,\hdots,1)$. With the resulting norm $k
= \sqrt{g_{\iota\iota}}$ of the Killing vector, a more
convenient expression for the metric (\ref{InitialMetric}) in
view of our later application, is given in terms of an
anholonomic vielbein co-base by
\begin{equation}
   ds^2 =   \eta_{ab} e^a e^b
          + e^i e^i
        =   \eta_{mn} e^m e^n \; ,
        \label{Original}
\end{equation}
where the chosen anholonomic vielbein $e^l = (e^a, e^i)$ reads
\begin{equation}
   e^a = e^a_{\phantom{a}\alpha} (x^\beta) dx^\alpha \; , \qquad\qquad
   e^i = e^i_{\phantom{i}\iota} dx^\iota
       = \delta^i_\iota k(x^\alpha) dx^\iota\; ,
\end{equation}
such that its components $e^l_{\phantom{l}\lambda}$ are
\begin{alignat}{3}
   &e^a_{\phantom{a}\alpha} = e^a_{\phantom{a}\alpha}(x^\beta) \; ,
   \qquad\qquad
   &&e^i_{\phantom{i}\alpha} = 0 \\
   &e^a_{\phantom{a}\iota} = 0 \; ,
   &&e^i_{\phantom{i}\iota} = \delta^i_\iota k \; .
      \label{InitialVielbein}
\end{alignat}
Later, we will also need the inverted vielbein of the isometry
direction $e_i^{\phantom{i}\iota}
= \delta_i^\iota/k$. \\
\phantom{a}\hspace{1em} The above mentioned $\sigma$-model action is
invariant
\cite{Verlinde} to first order in $\epsilon$
under the translational isometry $\delta_\epsilon x^\mu =
\epsilon k^\mu$, if $k^\mu$ satisfies the Killing equation
${({\cal L}_k g)}_{\mu\nu} = k^\lambda g_{\mu\nu,\lambda}
+ k^\lambda_{\phantom{\lambda},\mu} g_{\lambda\nu}
+ k^\lambda_{\phantom{\lambda},\nu} g_{\lambda\mu}
= \partial g_{\mu\nu}/ \partial x^\iota = 0$ and the torsion
obeys ${\cal L}_k H = 0$. This implies ${\cal L}_k B = dw$ for
some 1-form $w$. Here ${\cal L}_k$ and $d$ denote the space-time
Lie- and exterior derivative. Locally, this is solved by $w =
i_k B - v$ with $dv = -i_k H$ for some 1-form $v$. Under $i_k$
we understand the
 interior
product. From now on, we will choose the gauge $w=0$, which
leads us, together with the identity $ \left( i_k B \right)_\mu
= k^\nu B_{\nu\mu} = B_{\iota\mu}$, to
\begin{equation*}
   v_\alpha = B_{\iota\alpha} \; .
\end{equation*}
The T-dual metric ${\tilde g}_{\mu\nu}$ reduces in the case of
$k^\alpha=0$ to
\begin{equation*}
   {\tilde g}_{\iota\iota} = \frac{1}{k^2} \; , \qquad\quad
   {\tilde g}_{\alpha\iota} = \frac{v_\alpha}{k^2} \; , \qquad\quad
   {\tilde g}_{\alpha\beta} = g_{\alpha\beta}+\frac{v_\alpha v_\beta}{k^2}
       \; .
\end{equation*}
Thus the dual line-element can be written as
\begin{equation*}
   d{\tilde s}^2 = {\tilde g}_{\alpha\beta} dx^\alpha dx^\beta
                   +2{\tilde g}_{\alpha\iota} dx^\alpha dx^\iota
                   +{\tilde g}_{\iota\iota} dx^\iota
                    dx^\iota
                 = \eta_{ab} {\tilde e}^a {\tilde e}^b
                    + {\tilde e}^i {\tilde e}^i
                 = \eta_{mn} {\tilde e}^m {\tilde e}^n \; .
\end{equation*}
Here the T-dual anholonomic vielbein co-base ${\tilde e}^l =
({\tilde e}^a,{\tilde e}^i)$ is given by
\begin{equation*}
    {\tilde e}^a = e^a = e^a_{\phantom{a}\alpha} dx^\alpha \; , \qquad\qquad
   {\tilde e}^i = {\tilde e}^i_{\phantom{i}\mu} dx^\mu
                = \frac{\delta^i_\iota}{k} dx^\iota
                 +\frac{v_\alpha}{k} dx^\alpha
\end{equation*}
from which we can read off the dual vielbein components
${\tilde e}^l_{\phantom{l}\lambda}$
\begin{alignat}{3}
   &{\tilde e}^a_{\phantom{a}\alpha} = e^a_{\phantom{a}\alpha}
    \; , \qquad\qquad
   &&{\tilde e}^i_{\phantom{i}\alpha} = \frac{v_\alpha}{k} \\
   &{\tilde e}^a_{\phantom{a}\iota} = 0 \; ,
   &&{\tilde e}^i_{\phantom{i}\iota} = \frac{\delta^i_\iota}{k} \; .
      \label{DualVielbein}
\end{alignat}
and their inverses ${\tilde e}_l^{\phantom{l}\lambda}$
\begin{alignat}{3}
   &{\tilde e}_a^{\phantom{a}\alpha} = e_a^{\phantom{a}\alpha}
    \; , \qquad\qquad
   &&{\tilde e}_i^{\phantom{i}\alpha} = 0 \\
   &{\tilde e}_a^{\phantom{a}\iota} = -e_a^{\phantom{a}\alpha} v_\alpha \; ,
   &&{\tilde e}_i^{\phantom{i}\iota} = \delta_i^\iota  k \; .
      \label{DualVielbein}
\end{alignat}
A useful formula consists of the inverted relation
$\delta^i_\iota dx^\iota = k {\tilde e}^i -
e_a^{\phantom{a}\alpha} v_\alpha e^a$. From ${\tilde
g}^{\mu\nu} = {\tilde e}_m^{\phantom{m}\mu} {\tilde
e}_n^{\phantom{n}\nu} \eta^{mn}$ we derive the inverse metric
to be
\begin{equation*}
   {\tilde g}^{\alpha\beta} = g^{\alpha\beta} \; , \qquad\quad
   {\tilde g}^{\alpha\iota} = -g^{\alpha\beta} v_\beta \; , \qquad\quad
   {\tilde g}^{\iota\iota} = g^{\alpha\beta} v_\alpha v_\beta + k^2
       \; .
\end{equation*}
\phantom{a}\hspace{1em}
The strategy of the calculation of the dual Riemann-tensor will
now be as follows. With the aid of Cartan's structure
equations, we determine the curvature tensor and its T-dual in
the $e^l = \{e^a, e^i\}$, resp. T-dual ${\tilde e}^l = \{
{\tilde e}^a, {\tilde e}^i \}$ co-base. In order to compare
both of them, it is further necessary to switch to the
equivalent expressions in the common holonomic co-base $dx^\mu
= (dx^\alpha, dx^\iota)$ with the help of the aforementioned
vielbeins. \\
\phantom{a}\hspace{1em}
Therefore, let us begin with (\ref{Original}) and avail
ourselves of Cartan's first structure equation for the
torsion-less case, $de^m+\omega^m_{\phantom{m}n}\wedge e^n=0$,
to determine the connection
 1-form
$\omega$
\begin{equation*}
   \omega^a_{\phantom{a}b} = -\partial_{[\gamma} e^a_{\phantom{a}\alpha]}
                               e_{[c}^{\phantom{c}\gamma}
                               e_{b]}^{\phantom{b}\alpha}
                               e^c
                               \; , \qquad\quad
   \omega^i_{\phantom{i}a} = e_a^{\phantom{a}\alpha}
                              \partial_\alpha \ln k \cdot e^i
                               \; , \qquad\quad
   \omega^i_{\phantom{i}i} = 0 \; ,
\end{equation*}
Cartan's second structure equation, $d\omega^m_{\phantom{m}n}
+\omega^m_{\phantom{m}l} \wedge \omega^l_{\phantom{l}n}
= R^m_{\phantom{m}n}$, together with the expression for the
curvature 2-form in the initial anholonomic co-base,
$R^l_{\phantom{l}m}=\frac{1}{2} R^l_{\phantom{l}mnp} e^n \wedge
e^p = \frac{1}{2}R^l_{\phantom{l}mab} e^a \wedge e^b +
R^l_{\phantom{l}mia} e^i \wedge e^a$, allow us to extract after
some algebra the initial curvature tensor\footnote{We use the
following anti- and symmetrization convention:
        \begin{equation*}
           A_{(ab)} = \frac{1}{2}\left( A_{ab} + A_{ba} \right)
           \qquad\qquad
           A_{[ab]} = \frac{1}{2}\left( A_{ab} - A_{ba} \right)
        \end{equation*}}
as
\begin{alignat*}{3}
   R_{abcd} &= e_{[c}^{\phantom{[c}\gamma}
               e_{d]}^{\phantom{d]}\delta}
        \left( e_b^{\phantom{b}\beta} \partial_\beta \partial_\gamma
               e_{a\delta}
              -\partial_{[\delta}
               e_{|a|\beta]} \cdot
               \partial_\gamma e_b^{\phantom{b}\beta}
        \right)
              +e_b^{\phantom{b}\beta}
               e_{[c}^{\phantom{[c}\gamma} \partial_{|\beta|}
               e_{d]}^{\phantom{d]}\alpha} \cdot
               \partial_{[\gamma} e_{|a|\alpha]} \\
   R_{iabc} &= R_{abic} = 0 \\
   R_{iaib} &= -e_{(a}^{\phantom{(a}\alpha} \partial_{|\alpha|}
                e_{b)}^{\phantom{b)}\beta}
                \cdot \partial_\beta \ln k
               -e_a^{\phantom{a}\alpha}
                e_b^{\phantom{b}\beta}
         \left( \partial_\alpha \partial_\beta \ln k
               +\partial_\alpha \ln k \cdot
                \partial_\beta \ln k
         \right)  \; ,
\end{alignat*}
whereas all other components are zero. By multiplying with the
vielbein $R_{\lambda\mu\nu\pi} = e^l_{\phantom{l}\lambda}
e^m_{\phantom{m}\mu}$ $e^n_{\phantom{n}\nu}
e^p_{\phantom{p}\pi} R_{lmnp}$, we subsequently arrive
 at
the coordinate base expressions
\begin{alignat*}{3}
   R_{\alpha\beta\gamma\delta} &=
      \delta_{[\gamma}^\varepsilon e^b_{\phantom{b}\delta]}
      \left( e_{a\alpha} \partial_\beta e_b^{\phantom{b}\zeta} \cdot
             \partial_{[\varepsilon} e^a_{\phantom{a}\zeta]}
            +\frac{1}{2} e_a^{\phantom{a}\zeta} \partial_\zeta e_{b\alpha}
             \cdot \partial_\varepsilon e^a_{\phantom{a}\beta}
      \right)
     +e^a_{\phantom{a}\alpha}
      \partial_\beta \partial_{[\gamma}
      e_{|a|\delta]}
     +\frac{1}{2} \partial_{[\gamma} e^a_{\phantom{a}|\alpha|} \cdot
      \partial_{\delta]} e_{a\beta} \\
   R_{\iota\alpha\beta\gamma} &= R_{\alpha\beta\iota\gamma} = 0 \\
   R_{\iota\alpha\iota\beta} &= -k^2
          \left( e^a_{\phantom{a}(\alpha} \partial_{\beta)}
                 e_a^{\phantom{a}\varepsilon} \cdot
                 \partial_\varepsilon \ln k
                +\partial_\alpha \partial_\beta \ln k
                +\partial_\alpha \ln k \cdot
                 \partial_\beta \ln k
          \right) \; ,
\end{alignat*}
and all other components zero. \\
\phantom{a}\hspace{1em}
Analogously, application of $d{\tilde e}^l + {\tilde
\omega}^l_{\phantom{l}m} \wedge {\tilde e}^m$ yields
\begin{equation*}
   {\tilde \omega}^a_{\phantom{a}b} = \omega^a_{\phantom{a}b}
                                           \; , \qquad\quad
   {\tilde \omega}^i_{\phantom{i}a} = -e_b^{\phantom{b}\beta}
   e_a^{\phantom{a}\alpha} \frac{dv_{\beta\alpha}}{k} \; e^b
   -e_{a}^{\phantom{a}\alpha} \partial_\alpha \ln k \; {\tilde e}^i
                                           \; , \qquad\quad
   {\tilde \omega}^i_{\phantom{i}i} = 0 \; ,
\end{equation*}
where $dv_{\alpha\beta} \equiv \partial_{[\alpha} v_{\beta]}$
denotes the exterior derivative of $v$. From the relation
$dv=-i_k H$, it is possible to substitute $dv_{\alpha\beta}$ in
the subsequent formulae by the torsion 3-form through
$dv_{\alpha\beta} = -H_{\iota\alpha\beta}$. Again, from
Cartan's second structure equation, $d{\tilde
\omega}^m_{\phantom{m}n}
+{\tilde \omega}^m_{\phantom{m}l} \wedge {\tilde \omega}^l_{\phantom{l}n}
= {\tilde R}^m_{\phantom{m}n}$, and the dual curvature 2-form
${\tilde R}^l_{\phantom{l}m} =
\frac{1}{2} {\tilde R}^l_{\phantom{l}mnp} {\tilde e}^n \wedge {\tilde e}^p =
\frac{1}{2} {\tilde R}^l_{\phantom{l}mab} e^a \wedge e^b
+ {\tilde R}^l_{\phantom{l}mia} {\tilde e}^i \wedge e^a$, we get the
following non-coordinate frame expressions
\begin{alignat*}{3}
   {\tilde R}_{abcd} &= R_{abcd}-\frac{2}{k^2}
                        e_a^{\phantom{a}\alpha}
                        e_b^{\phantom{b}\beta}
                        e_{[c}^{\phantom{[c}\gamma}
                        e_{d]}^{\phantom{d]}\delta}
                        dv_{\alpha\gamma}
                        dv_{\beta\delta} \\
   {\tilde R}_{iabc} &= {\tilde R}_{bcia}
                      =-\frac{2}{k} e_a^{\phantom{a}\alpha}
                        e_{[b}^{\phantom{[b}\beta}
                        e_{c]}^{\phantom{c]}\gamma}
                        dv_{\beta\alpha} \cdot
                        \partial_\gamma \ln k \\
   {\tilde R}_{iaib} &= e_{(a}^{\phantom{(a}\alpha}
                        \partial_{|\alpha|}
                        e_{b)}^{\phantom{b)}\beta} \cdot
                        \partial_\beta \ln k
                       +e_a^{\phantom{a}\alpha}
                        e_b^{\phantom{b}\beta}
                 \left( \partial_\alpha \partial_\beta \ln k
                       -\partial_\alpha \ln k \cdot
                        \partial_\beta \ln k
                 \right) \; .
\end{alignat*}
Switching to the coordinate base with the help of ${\tilde
R}_{\lambda\mu\nu\pi} = {\tilde e}^l_{\phantom{l}\lambda}
{\tilde e}^m_{\phantom{m}\mu} {\tilde e}^n_{\phantom{n}\nu}
{\tilde e}^p_{\phantom{p}\pi} {\tilde R}_{lmnp}$, one finally
arrives, after some tedious algebra, at the desired
expressions, which relate the T-dual curvature tensor to its
counterpart for the initial metric
\begin{alignat}{3}
    {\tilde R}_{\alpha\beta\gamma\delta} &= R_{\alpha\beta\gamma\delta}
       +\frac{2}{k^2}
        \bigg( dv_{\alpha[\delta} dv_{|\beta|\gamma]}
                     -dv_{\alpha\gamma} \cdot
                      v_{[\beta} \partial_{\delta]} \ln k
                     +dv_{\beta\gamma} \cdot
                      v_{[\alpha} \partial_{\delta]} \ln k
                     +dv_{\alpha\delta} \cdot
                      v_{[\beta} \partial_{\gamma]} \ln k    \notag \\
       &\qquad\qquad\qquad
                     -dv_{\beta\delta} \cdot
                      v_{[\alpha} \partial_{\gamma]} \ln k
                     +v_{[\alpha} e^b_{\phantom{b}\beta]}
                      v_{[\gamma} \partial_{\delta]}
                      e_b^{\phantom{b}\varepsilon} \cdot
                      \partial_\varepsilon \ln k
                     +v_{[\gamma} e^b_{\phantom{b}\delta]}
                      v_{[\alpha} \partial_{\beta]}
                      e_b^{\phantom{b}\varepsilon} \cdot
                      \partial_\varepsilon \ln k             \notag \\
       &\qquad\qquad\qquad
                     +v_\gamma v_{[\alpha} \partial_{\beta]}
                      \partial_\delta \ln k
                     -v_\delta v_{[\alpha} \partial_{\beta]}
                      \partial_\gamma \ln k
              -2 v_{[\alpha} \partial_{\beta]} \ln k \cdot
                 v_{[\gamma} \partial_{\delta]} \ln k
        \bigg)
          \label{Riemann1} \\
    {\tilde R}_{\iota\alpha\beta\gamma} &= \frac{2}{k^2}
        \bigg( dv_{\alpha[\beta} \cdot \partial_{\gamma]} \ln k
              +\frac{1}{2}\big( e^a_{\phantom{a}\alpha} v_{[\beta}
                                \partial_{\gamma]} e_a^{\phantom{a}\delta}
                               +v_{[\beta} e^a_{\phantom{a}\gamma]}
                                \partial_\alpha e_a^{\phantom{a}\delta}
                          \big) \cdot \partial_\delta \ln k
              +v_{[\beta} \partial_{\gamma]} \partial_\alpha \ln k
                                                              \notag \\
      &\qquad\;
              -v_{[\beta} \partial_{\gamma]} \ln k \cdot \partial_\alpha
               \ln k
        \bigg)
          \label{Riemann2} \\
    {\tilde R}_{\iota\alpha\iota\beta} &=
               -\frac{1}{k^2}
                \bigg( \frac{R_{\iota\alpha\iota\beta}}{k^2}
                      +2 \partial_\alpha \ln k \cdot
                         \partial_\beta \ln k
                \bigg) \; .
          \label{Riemann3}
\end{alignat}
\phantom{a} \\
\phantom{a}\hspace{1em}
As an application of these formulae, we now want to examine,
whether d-dimensional $\ads_d$ space allows a T-dualization to
flat space under the inclusion of some suitably chosen
$B_{\mu\nu}$. Because both spaces, regarded as string
backgrounds, leave all supersymmetries intact this is a
reasonable question to ask. We choose the following coordinate
representation for $\ads_d$
\begin{equation}
   ds^2 = \frac{r^2}{L^2} \bigg( -dt^2 + \sum_{j=1}^{d-2} dy^j dy^j \bigg)
         + L^2 \frac{dr^2}{r^2} \; ,
\end{equation}
with $L$ being the AdS-radius. It exhibits a negative, constant
curvature scalar $R=-d(d-1)/L^2$. Let us first choose for the
isometry direction one of the $y^j$ coordinates,
e.g.~$y^{d-2}$. We will see below how the results for this case
extrapolate to the case of a different isometry direction. In
our above convention, the adapted coordinates are $x^0 = t,\,
x^1 = y^1, \hdots , x^{d-3} = y^{d-3}, \, x^{d-2} = r, \,
x^{d-1} \equiv x^\iota = y^{d-2}$. The curvature tensor of the
negatively curved, maximally symmetric $\ads_d$ is given by
$R_{\lambda\mu\nu\pi} = - \frac{1}{L^2}
                          \left( g_{\lambda\nu} g_{\mu\pi}
                                -g_{\lambda\pi} g_{\mu\nu}
                          \right)$,
which reduces in our parameterization to\footnote{We choose the
Minkowski
                                                  metric to be $\eta=
                                                  (-,+,\hdots,+)$.}
\begin{equation*}
   R_{\mu\nu\mu\nu} = -\frac{r^4}{L^6} \eta_{\mu\mu}\eta_{\nu\nu}  \; ,
   \qquad\qquad
   R_{\mu r \mu r} = -\frac{1}{L^2}\eta_{\mu\mu} \; ,
   \qquad  \mu,\nu \not= r
\end{equation*}
and all other components vanishing. In particular, this gives
$R_{\iota\alpha\iota\beta} = -\frac{r^4}{L^6}\eta_{\alpha\beta}
+\big(\frac{r^4}{L^6}-\frac{1}{L^2}\big) \delta^r_\alpha \delta^r_\beta$.
Now, we had found one component, (\ref{Riemann3}), of the dual
curvature-tensor, which was independent of $v$, resp.
$B_{\mu\nu}$ and hence allows answering the above posed
question. Plugging in the actual norm of the Killing vector
$k=r/L$ for the above $\ads_d$ metric, we obtain for the
right-hand side of (\ref{Riemann3}) the expression
$\frac{1}{L^2}\eta_{\alpha\beta}-\big(\frac{1}{L^2}+\frac{L^2}{r^4}\big)
\delta^r_\alpha \delta^r_\beta$. Obviously, this does not fulfill the
requirement ${\tilde R}_{\iota\alpha\iota\beta}=0$ for flat
space and is not even asymptotically flat. This can be
extrapolated to any other isometry direction of $\ads_d$ by
noting that another isometry direction different from the above
choice of $y^{d-2}$ can be related to this one by a general
coordiante transformation $x\rightarrow x'$. However, the tensor
transformation property ${{\tilde
R}'}_{\iota'\alpha'\iota'\beta'} (x')  =
\left( A^\iota_{\iota'} \right)^2 A^\alpha_{\alpha'} A^\beta_{\beta'}
{\tilde R}_{\iota\alpha\iota\beta} (x)$, where $A^\mu_\nu=
\partial x^\mu /
\partial {x'}^{\nu}$, then guarantees that this term is still
non-vanishing in contrast to what one would need for a dual
flat spacetime. Thus, we can conclude that flat space cannot be
reached from pure $\ads_d$
via T-duality under the inclusion of some appropriately chosen $B_{\mu\nu}$.\\
\phantom{a}\hspace{1em}
Eventually, we turn briefly to the important $\ads_5 \times
\text{S}^5$ extension of $\ads$. Here it seems possible to
relate the IIB D9-brane geometry, which describes D=10 space
itself, to the $\ads_5 \times \text{S}^5$ geometry. In order to
do so, one has to start with a six-fold T-duality, transforming
the D9-brane to the D3-brane. Then, dualizing a IIB D3-brane to
its own near-horizon geometry, one reaches $\ads_5 \times S^5$.
This can be done following the work of
\cite{Skenderis}.
Having T-dualized the D9-brane, the D3-brane geometry in
string-frame reads
\begin{alignat*}{3}
   ds^2 &= \frac{1}{\sqrt{H_6}}
           \big( -dt^2 + \sum_{i=1}^3 (dx^i)^2 \big)
          +\sqrt{H_6}
           \sum_{j=4}^9 (dx^j)^2 \; , \\
   &\;e^\phi = 1 \; ,  \qquad\qquad\qquad\quad\;
   C_{0123} = \frac{1}{H_6}-1 \; , \\
   &H_6 = 1+\frac{Q_{D3}}{(r_6)^4} \; ,  \qquad\qquad
   Q_{D3} = 4\pi g_s N l_s^4 \; .
\end{alignat*}
Performing a further T-duality over $x^3, x^2$ brings us to the
IIB D1-brane. An S-duality transformation then yields the
fundamental IIB string solution
\begin{alignat}{3}
   ds^2 &= \frac{1}{H_8}
           \big( -dt^2 + (dx^1)^2 \big)
          + \sum_{j=2}^9 (dx^j)^2 \; , \\
   e^\phi &= \frac{1}{\sqrt{H_8}} \; ,  \qquad\qquad\quad\;
   B_{01} = \frac{1}{H_8}-1 \; , \\
   H_8 &= 1+\frac{Q_{F1}}{(r_8)^6} \; , \qquad\qquad\!
   Q_{F1} = d_1 g_s^2 N l_s^6 \; ,
\end{alignat}
which is then T-dualized over $x_1$ to obtain the IIA
gravitational wave solution
\begin{alignat}{3}
   ds^2 &=  \left( H_8-2 \right) dt^2
           + 2 \left( H_8-1 \right) dt dx^1
           + H_8 dx^1 dx^1
           + \sum_{j=2}^9 (dx^j)^2 \; , \\
   &\qquad\qquad\qquad
   e^\phi = 1 \; ,  \qquad\qquad
   B_{\mu\nu} = 0 \; .
               \label{GravWave}
\end{alignat}
The crucial step to get rid of the constant in the harmonic
$H_8$ function, consists of performing first an
$SL(2,\mathbb{R})$ coordinate transformation on the coordinates
$t, x^1$ and subsequently a T-duality transformation in the new
${x'}^1$ direction. Provided that we choose the special
$SL(2,\mathbb{R})$ coordinate transformation
\begin{equation}
   t' = \frac{1}{2}\left( t + x^1 \right) \; , \qquad
   {x'}^1 = 2 x^1 \; ,
\end{equation}
we are guaranteed that the transformation is globally
well-defined on $(t, x^1)$ space, which has the topology of a
cylinder due to the compactification of $x^1$. After T-duality
in ${x'}^1$ direction the result is a modified fundamental
string without constant part in the harmonic function
\begin{alignat}{3}
   ds^2 &= \frac{1}{{\cal H}_8}
           \big( -(dt')^2 + (d{x'}^{1})^2 \big)
          + \sum_{j=2}^9 (dx^j)^2 \; , \\
   e^\phi &= \frac{1}{\sqrt{{\cal H}_8}} \; , \qquad\qquad\quad\;
   B_{01} = \frac{1}{{\cal H}_8}-2 \; , \\
   &\qquad\qquad\; {\cal H}_8 = \frac{Q_{F1}}{(r_8)^6} \;
\end{alignat}
It is important to note that the last T-duality has again been
along a space-like direction. Therefore we are secured to stay
in the Type II string-theory framework instead of changing to
Type II* \cite{Hull}. \\
\phantom{a}\hspace{1em}
Now, proceeding in the inverse manner, a second S-duality
promotes us to a modified D1-brane. Ultimately, with two
T-dualities in the $x^2, x^3$ directions we end up with a
modified D3-brane solution without a constant part in its
harmonic function. As is well-known, this gives the $\ads_5
\times \text{S}^5$ geometry. Notice, however, that the D3-brane
solution is only locally dual to the $\ads_5 \times \text{S}^5$
solution. In order to perform the various T-dualities, the
brane has to be wrapped on a torus with all worldvolume
coordinates taken to be periodic. Therefore the coordinates of
the final $\ads_5 \times \text{S}^5$ geometry in addition have
to be identified globally. As mentioned in \cite{Skenderis}
this has the effect that only half the Killing spinors of
$\ads_5$ remain. Thus, only locally we have a maximally
supersymmetric $\ads_5 \times \text{S}^5$ solution. Globally,
the
D3-brane as well as the dual solution break half the supersymmetry. \\
\phantom{a}\hspace{1em}
A word of caution is in order. The spacetime-filling D9-brane
is supposed to describe flat ten-dimensional Minkowski spacetime
\cite{deRoo}. However, the D9-brane breaks half the supersymmetry,
whereas flat space does not. The resolution to this puzzle comes
from the well-known fact, that the open-string sector of $N$
D9-branes is only consistent, if we have $N=32$ of them and
perform an orientifolding by the world-sheet parity $\Omega:\;
\sigma \rightarrow \pi-\sigma$. This breaks half the
supersymmetry and the resulting Type I SO(32) theory gives us
in addition to the flat ten-dimensional Minkowski solution a
spacetime-filling O9 orientifold fixed plane. It would be
interesting to study the influence of dualities on the
cosmological constant in string-theory more generally. In
particular the strongly coupled heterotic string \cite{HW} whose
non-perturbative open membrane boundary-boundary interactions
lead to a non-vanishing positive vacuum energy \cite{CK2} (see
also \cite{K1}) would be an interesting candidate to study.

\bigskip
\noindent {\large \bf Acknowledgements}\\[2ex]
Financial support by the DFG is gratefully acknowledged.

\newpage

 \newcommand{\zpc}[3]{{\sl Z. Phys.} {\bf C#1} (#2) #3}
 \newcommand{\zp}[3]{{\sl Z. Phys.} {\bf #1} (#2) #3}
 \newcommand{\npb}[3]{{\sl Nucl. Phys.} {\bf B#1} (#2)~#3}
 \newcommand{\plb}[3]{{\sl Phys. Lett.} {\bf B#1} (#2) #3}
 \newcommand{\prd}[3]{{\sl Phys. Rev.} {\bf D#1} (#2) #3}
 \newcommand{\prl}[3]{{\sl Phys. Rev. Lett.} {\bf #1} (#2) #3}
 \newcommand{\prep}[3]{{\sl Phys. Rep.} {\bf #1} (#2) #3}
 \newcommand{\fp}[3]{{\sl Fortsch. Phys.} {\bf #1} (#2) #3}
 \newcommand{\nc}[3]{{\sl Nuovo Cimento} {\bf #1} (#2) #3}
 \newcommand{\ijmp}[3]{{\sl Int. J. Mod. Phys.} {\bf #1} (#2) #3}
 \newcommand{\rmp}[3]{{\sl Rev. Mod. Phys.} {\bf #1} (#2) #3}
 \newcommand{\ptp}[3]{{\sl Prog. Theor. Phys.} {\bf #1} (#2) #3}
 \newcommand{\sjnp}[3]{{\sl Sov. J. Nucl. Phys.} {\bf #1} (#2) #3}
 \newcommand{\cpc}[3]{{\sl Comp. Phys. Commun.} {\bf #1} (#2) #3}
 \newcommand{\mpla}[3]{{\sl Mod. Phys. Lett.} {\bf A#1} (#2) #3}
 \newcommand{\cmp}[3]{{\sl Commun. Math. Phys.} {\bf #1} (#2) #3}
 \newcommand{\jmp}[3]{{\sl J. Math. Phys.} {\bf #1} (#2) #3}
 \newcommand{\nim}[3]{{\sl Nucl. Instr. Meth.} {\bf #1} (#2) #3}
 \newcommand{\el}[3]{{\sl Europhysics Letters} {\bf #1} (#2) #3}
 \newcommand{\ap}[3]{{\sl Ann. of Phys.} {\bf #1} (#2) #3}
 \newcommand{\atmp}[3]{{\sl Adv. Theor. Math. Phys.} {\bf #1} (#2) #3}
 \newcommand{\cqg}[3]{{\sl Class. Quant. Grav.} {\bf #1} (#2) #3}
 \newcommand{\jetp}[3]{{\sl JETP} {\bf #1} (#2) #3}
 \newcommand{\jetpl}[3]{{\sl JETP Lett.} {\bf #1} (#2) #3}
 \newcommand{\jhep}[3]{{\sl JHEP} {\bf #1} (#2) #3}
 \newcommand{\acpp}[3]{{\sl Acta Physica Polonica} {\bf #1} (#2) #3}
 \newcommand{\vj}[4]{{\sl #1~}{\bf #2} (#3) #4}
 \newcommand{\ej}[3]{{\bf #1} (#2) #3}
 \newcommand{\vjs}[2]{{\sl #1~}{\bf #2}}
 \newcommand{\hep}[1]{{\sl hep--ph/}{#1}}
 \newcommand{\desy}[1]{{\sl DESY-Report~}{#1}}


\begin{thebibliography}{99}

  \bibitem{Malda} J.~Maldacena, {\it The large N Limit of Superconformal
                  Field Theories and Supergravity}, \atmp{2}{1998}{231},
                  hep-th/9711200;

  \bibitem{Gubser} S.S.~Gubser, I.R.~Klebanov and A.M.~Polyakov,
                   {\it Gauge Theory Correlators from Noncritical
                   String Theory}, \plb{428}{98}{105}, hep-th/9802109;

  \bibitem{Witten} E.~Witten, {\it Anti-de Sitter Space and Holography},
                   \atmp{2}{1998}{253}, hep-th/9802150;

  \bibitem{Malda2} O.~Aharony, S.S.~Gubser, J.~Maldacena, H.~Ooguri
                   and Y.~Oz, {\it Large N Field Theories, String Theory
                   and Gravity}, hep-th/9905111;

  \bibitem{Polyakov} D.~Polyakov, {\it On the NSR Formulation of String
                     Theory on $AdS_5\times S^5$}, hep-th/9812044;

  \bibitem{Tseytlin} R.R.~Metsaev and A.A.~Tseytlin, {\it Supersymmetric
                     D3-brane Action in $AdS_5\times S^5$},
                     \plb{436}{1998}{281}, hep-th/9806095;
                     R.~Kallosh and A.A.~Tseytlin, {\it Simplifying
                     Superstring Action on $AdS_5\times S^5$},
                     \jhep{9810}{1998}{016}, hep-th/9808088;
                     I.~Pesando, {\it A Kappa fixed Type IIB Superstring
                     Action on $AdS_5\times S^5$},
                     \jhep{9811}{1998}{002}, hep-th/9808020;
                     R.~Kallosh, J.~Rahmfeld and A.~Rajaraman,
                     {\it Near Horizon Superspace},
                     \jhep{9809}{1998}{002}, hep-th/9805217;
                     R.~Kallosh and J.~Rahmfeld,
                     {\it The GS String Action on $AdS_5\times S^5$},
                     \plb{443}{1998}{143}, hep-th/9808038;

  \bibitem{K3} E.~Verlinde, {\it On RG Flow and the Cosmological
               Constant}, \cqg{17}{2000}{1277}, hep-th/9912058;
               C.~Schmidhuber, {\it ADS(5) and the 4D
               Cosmological Constant}, \npb{580}{2000}{140},
               hep-th/9912156;
               A.~Krause, {\it A Small Cosmological Constant,
               Grand Unification and Warped Geometry},
               hep-th/0006226;
               {\it A Small Cosmological Constant and
               Backreaction of Non-Finetuned Parameters},
               \jhep{0309}{2003}{016}, hep-th/0007233;
               {\it Heterotic M-Theory, Warped Geometry and the
               Cosmological Constant Problem},
               \fp{49}{2001}{163};

  \bibitem{Alvarez} E.~Alvarez, L.~Alvarez-Gaume, J.L.F.~Barbon and
                    Y. Lozano,
                    {\it Some Global Aspects of Duality in String Theory},
                    \npb{415}{1994}{71}, hep-th/9309039;

  \bibitem{Verlinde} M.~Ro\v{c}ek and E.~Verlinde,
                     {\it Duality, Quotients and Currents},
                     \npb{373}{1992}{630}, hep-th/9110053;

  \bibitem{Skenderis} H.J.~Boonstra, B.~Peeters and K.~Skenderis,
                      {\it Duality and Asymptotic Geometries},
                      \plb{411}{1997}{59}, hep-th/9706192;

  \bibitem{Hull} C.~M.~Hull,
                 {\it Timelike T-Duality, de Sitter Space, Large N
                 Gauge Theories and Topological Field Theory},
                 \jhep{9807}{1998}{021}, hep-th/9806146;

  \bibitem{deRoo} E.~Bergshoeff and M.~de Roo, {\it D-Branes and
                  T-Duality}, \plb{380}{1996}{265},
                  hep-th/9603123;

  \bibitem{HW} P.~Ho\v{r}ava and E.~Witten, {\it Heterotic and Type I
               String Dynamics from Eleven Dimensions}, \npb{460}{1996}{506},
               hep-th/9510209; {\it Eleven-Dimensional Supergravity on a
               Manifold with Boundary}, \npb{475}{1996}{94},
               hep-th/9603142; E.~Witten, {\it Strong Coupling Expansion of
               Calabi-Yau Compactification}, \npb{471}{1996}{135},
               hep-th/9602070; A.~Lukas, B.A.~Ovrut and D.~Waldram,
               {\it On the Four-Dimensional Effective Action of Strongly
               Coupled Heterotic String Theory}, \npb{532}{1998}{43},
               hep-th/9710208; {\it The Ten-Dimensional Effective Action
               of Strongly Coupled Heterotic String Theory},
               \npb{540}{1999}{230}, hep-th/9801087; G.~Curio and A.~Krause,
               {\it Four-Flux and Warped Heterotic M-Theory Compactifications},
               \npb{602}{2001}{172}, hep-th/0012152;

  \bibitem{CK2} G.~Moore, G.~Peradze and N.~Saulina,
                {\it Instabilities in Heterotic M-Theory induced by Open
                Membrane Instantons}, \npb{607}{2001}{117},
                hep-th/0012104;
                G.~Curio and A.~Krause, {\it G-Fluxes and Non-Perturbative
                Stabilisation of Heterotic M-Theory}, \npb{643}{2002}{131},
                hep-th/0108220;

  \bibitem{K1} C.~Bachas and B.~Pioline, {\it High-Energy
               Scattering on Distant Branes},
               \jhep{9912}{1999}{004}, hep-th/9909171;
               A.~Krause, {\it Testing Stability of M-Theory on
               an $S^1/Z_2$ Orbifold}, \jhep{0005}{2000}{046},
               hep-th/9909182;
               T.~Dasgupta, M.R.~Gaberdiel and M.B.~Green, {\it
               The Type I D-Instanton and its M-Theory Origin},
               \jhep{0008}{2000}{004}, hep-th/0005211.

\end{thebibliography}
\end{document}